\documentclass[aps,prb,twocolumn,superscriptaddress]{revtex4-2}

\usepackage{xcolor}
\usepackage[
colorlinks=true,
citecolor=blue,
linkcolor=blue,
urlcolor=blue
]{hyperref}

\usepackage{amsmath}
\usepackage{amsthm}
\usepackage{amssymb}
\usepackage{amsfonts}
\usepackage{graphicx}
\usepackage{romannum}
\usepackage{float}
\usepackage{placeins}
\usepackage{soul}
\usepackage{appendix}

\begin{document}

	\title{Electric-field switchable interlayer magnetic order and anomalous valley Hall effect in Janus VSSe bilayers with different interfaces}

	\author{Yueli Li}
	
	\affiliation{School of Physics, Hefei University of Technology, Hefei 230009, People's Republic of China}
	
	\author{Yunfan Zhang}
	
	\affiliation{School of Physics, Hefei University of Technology, Hefei 230009, People's Republic of China}
	
	\author{Jiayu Dai}
	
	\affiliation{School of Physics, Hefei University of Technology, Hefei 230009, People's Republic of China}
	
	\author{Zhongjun Li}
	
	\affiliation{School of Physics, Hefei University of Technology, Hefei 230009, People's Republic of China}
	
	\author{Hongyan Lv}
	\email[Corresponding author: ]{hylv@hfut.edu.cn}
	\affiliation{School of Physics, Hefei University of Technology, Hefei 230009, People's Republic of China}


	\begin{abstract}
		Electric-field control of magnetic order and valley polarization holds great promise for spintronic and valleytronic applications. However, achieving such electrical modulation remains a fundamental challenge in two-dimensional (2D) van der Waals (vdW) magnets. Herein, via first-principles calculations, we verify that electrically tunable interlayer magnetic order and valley polarization can be realized in Janus VSSe bilayers with different interfaces. Although the magnitudes of dipole moments within each constituent Janus monolayer are identical, the vertical built-in electrostatic potential difference $\Delta\phi$ across the bilayer depends strongly on the orientations of these dipoles, which originates from the distinct interfacial configurations. As a consequence, the VSSe bilayers with different interfaces possess distinct interlayer magnetic couplings and show dramatically varied responses to external electric fields. For Se-S interface, reversible electric-field switching between antiferromagnetic (AFM) and ferromagnetic (FM) states can be achieved due to the competition between itinerant-electron-mediated FM exchange coupling and interlayer $\Delta\phi$-dependent AFM/FM super-superexchange interactions mediated by interfacial Se and S atoms. For Se-Se interface, both the valley polarization and spin splitting can be effectively reversed by the out-of-plane electric field, realizing an all-electric-field controlled anomalous valley Hall effect. Our results demonstrate that the interface of the magnetic Janus bilayer could act as an additional degree of freedom to tune the electronic, magnetic, and valley properties in 2D vdW materials. The Janus VSSe bilayers are identified as a promising platform for the design of low-power spintronic and valleytronic devices.

	\end{abstract}

	\maketitle
	\pagenumbering{arabic}
	
\section{INTRODUCTION}
	
Two-dimensional (2D) van der Waals (vdW) magnets have emerged as a promising platform for exploring low-dimensional magnetism and developing next-generation spintronic devices because of their unique layered structures and diverse magnetic phases \cite{1nature,2nature,3nat,4nat,5nature,6nature}. Although the Mermin-Wagner theorem predicts that long-range magnetic ordering cannot exist in isotropic 2D systems at finite temperatures \cite{7PhysRevLett.17.1133}, the experimental discovery of intrinsic ferromagnetism in $\mathrm{Cr_2Ge_2Te_6}$ and monolayer $\mathrm{CrI_3}$ demonstrated that magnetic anisotropy can stabilize long-range magnetic order in atomically thin layers \cite{1nature,2nature}. Since then, 2D magnetic materials have attracted extensive research interest and a variety of intrinsic 2D magnets have been reported \cite{3nat,4nat,5nature,6nature}. Owing to their reduced dimensionality and weak interlayer interactions, the electronic and magnetic properties of these materials can be effectively tuned by external stimuli, such as strain, carrier doping, and electric fields \cite{5nature,6nature}. The 2D magnetism also offers opportunities for coupling spin with other quantum degrees of freedom, especially the valley \cite{8PhysRevLett.99.236809,9nat}, thereby opening new possibilities for multifunctional spintronic and valleytronic devices.

The valley degree of freedom, which refers to a local minimum in the conduction band or local maximum in the valence band, provides an additional information carrier beyond charge and spin \cite{8PhysRevLett.99.236809,9nat}. In transition metal dichalcogenides (TMDs), 2$H$-phase MoS$_2$ monolayer for example, broken inversion symmetry combined with strong spin–orbit coupling (SOC) gives rise to valley-contrasting Berry curvature and spin-valley locking, enabling various valley-dependent optical and transport phenomena \cite{10PhysRevLett.108.196802,11xu2014spin}. To make use of the valley index, a material platform should ideally support controllable valley polarization. Nevertheless, the valley degeneracy protected by time-reversal symmetry in nonmagnetic systems prevents spontaneous valley polarization \cite{12RevModPhys.82.1959}. Introducing magnetism lifts the valley degeneracy, giving rise to ferrovalley states in which magnetic order and valley polarization coexist \cite{13tong2016}. The resulting Berry-curvature imbalance further induces anomalous transverse transport, such as the anomalous valley Hall (AVH) effect, highlighting the potential of ferrovalley materials for spintronic and valleytronic applications \cite{13tong2016,14LaBr2-APL-2019,15Nb3I8-PRB-2020}.

Efficient manipulation of magnetic order and valley polarization is of fundamental importance for realizing spintronic and valleytronic devices. For the tuning of magnetic order, various approaches, such as stacking \cite{16D.Xiao-NanoLett-2018,17C.Gao-Science-2019}, strain engineering \cite{18-1X.Xu-strain-2022,18Bilican-PRA-2026}, and carrier doping \cite{19jiang2018,20W.Shi-NatCom-2025}, have been explored in 2D magnetic materials. As for the manipulation of valley polarization and thus the AVH effect, previously reported strategies are primarily based on multiferroic materials, controlling valley properties through ferroelectricity, either via sliding ferroelectricity \cite{20-2Ren-PRL-2020,20-3Li-NanoLett-2024,20-4Ren-PRB-2025} or by forming heterostructures \cite{20-5Duan-JMCC-2020,20-6He-PRB-2021,20-7Ren-PRB-2022}. Compared with these methods, an external electric field offers a reversible, high-speed, and low-power route for tuning magnetic interactions, valley polarization, and valley-dependent transports \cite{21Mak-NatMat-2018,22huang2018,23Liu2025,24Kan-NatCom-2025,25Duan-npjQM-2017,26Ding-JPCL-2025}. However, successful electric-field control of interlayer magnetic order has been limited to only a few magnetic materials, including bilayers of CrI$_3$ and Cr$_2$Ge$_2$Te$_6$, and CrI$_3$/MnSe$_2$ heterobilayer \cite{21Mak-NatMat-2018,22huang2018,23Liu2025,24Kan-NatCom-2025}, and direct electric-field control of valley polarization remains rarely reported \cite{25Duan-npjQM-2017,26Ding-JPCL-2025}. 
    
Janus TMDs emerge as a new family of materials with broken out-of-plane mirror symmetry due to the different chalcogen layers, resulting in the intrinsic out-of-plane electric dipoles \cite{27lu2017,28dong2017large}. Among them, the Janus VSSe monolayer is a room-temperature ferromagnetic (FM) semiconductor that exhibits spontaneous valley polarization due to the coexistence of broken inversion symmetry, SOC, and magnetic order \cite{29-1Du-NanoLett-2019}. Owing to the additional breaking of mirror symmetry, its physical properties are more tunable by external electric fields. Previous studies have shown that both band structure and valley splitting of Janus VSSe monolayer can be continuously tuned by electric fields \cite{29-2Luo-PRB-2020}. Notably, the intrinsic dipole moment of Janus TMDs can be used to tune the interlayer interactions in Janus heterobilayers \cite{30zhang2021}. This inspires us to consider that when two Janus VSSe monolayers are stacked together, the relative orientations of intrinsic dipole moments within each constituent monolayer, which originates from the distinct interfaces, could serve as another degree of freedom to tune the associated properties. 

In this work, we perform first-principles calculations to systematically investigate the structural, electronic, magnetic, and valley properties of Janus VSSe bilayers with different interfaces. We demonstrate that the vertical built-in electrostatic potential difference depends strongly on the orientations of dipole moments within each constituent monolayer, arising from the distinct interfaces. Consequently, the interface of the Janus bilayer has a significant influence on the interlayer magnetic coupling. Electric-field controlled switching between antiferromagnetic (AFM) and FM states can be achieved for the Se-S interface. For the Se-Se interface, both the valley polarization and spin splitting can be effectively reversed by the out-of-plane electric field, realizing an all-electric-field controlled AVH effect. Our results demonstrate that the interface of the magnetic Janus bilayer could act as an additional degree of freedom to tune the electronic, magnetic, and valley properties in 2D vdW materials.

\section{COMPUTATIONAL DETAILS}
	
Our first-principles calculations were performed within the framework of density functional theory (DFT) as implemented in the Vienna $ab$ $initio$ simulations package (VASP) \cite{32vasp1,33vasp2}. The exchange-correlation functional was treated using the generalized gradient approximation (GGA) in the Perdew-Burke-Ernzerhof (PBE) form \cite{34PBE}. The core electrons were treated using the projector augmented wave method \cite{35PhysRevB.50.17953}. To account for the correlation effects of the V-$3d$ electrons, the DFT+$U$ method in the Dudarev formulation was employed \cite{36PhysRevB.57.1505}, with an effective Hubbard parameter $U_{\mathrm{eff}}$ of 2.0~eV. A plane-wave cutoff energy of 700~eV was used, and the Brillouin zone was sampled using an $18\times18\times1$ Monkhorst-Pack $k$-point mesh. A vacuum layer larger than 20~\AA{} was introduced along the out-of-plane direction to eliminate spurious interactions between periodic images. During structural optimization, vdW interactions were included using the optB86b-vdW functional \cite{37PhysRevLett.92.246401,38PhysRevB.83.195131}. The convergence criteria for the total energy and atomic forces were set to be $10^{-7}$~eV and $0.001$~eV/\AA{}, respectively. SOC was included in the calculations of the magnetic anisotropy and valley-related properties.
	
Tight-binding Hamiltonians were formulated in the basis of maximally localized Wannier functions (MLWFs) using the Wannier90 code \cite{39PhysRevB.56.12847}. Berry curvatures were evaluated with the WannierTools package \cite{40wanniertools}. The magnetic exchange parameters $J_{ij}$ were extracted based on the Green's function using TB2J package \cite{41tb2j,41-2green}. The spin Hamiltonian has the following form:
\[\displaystyle H = - \sum_{i \textless j}{J_{ij}\vec{S}_{i}\cdot\vec{S}_{j}} - \sum_{i}{A_i{S}_{iz}^{2}} - \sum_{i \textless j}{\vec{D}_{ij}\cdot(\vec{S}_i\times\vec{S}_j)},\]
where $J_{ij}$ represents the isotropic exchange interaction between magnetic atoms at sites $i$ and $j$, and $\vec{S}_{i}$ and $\vec{S}_{j}$ are the corresponding unit vectors denoting the directions of local magnetic moments. $A_{i}$ accounts for single-ion anisotropic energy and $\vec{D}_{ij}$ represents the Dzyaloshinshii-Moriya (DM) interaction. Magnetic transition temperatures were estimated by solving the atomistic Landau-Lifshitz-Gilbert (LLG) equation, as implemented in the VAMPIRE package \cite{42Evans_2014}. The simulations were carried out using the LLG-Heun integrator, with $6{,}000{,}000$ equilibration steps followed by $6{,}000{,}000$ averaging steps. An external electric field was applied along the out-of-plane direction with dipole corrections \cite{43PhysRevB.46.16067}. Phonon spectra were calculated by using $4\times4\times1$ supercells based on the density functional perturbation theory (DFPT) via the PHONOPY package \cite{44phonon}.

\section{RESULTS AND DISCUSSION}
	
\subsection{Structural properties}

\begin{figure*}
	\centering
	\includegraphics[width=1.6\columnwidth]{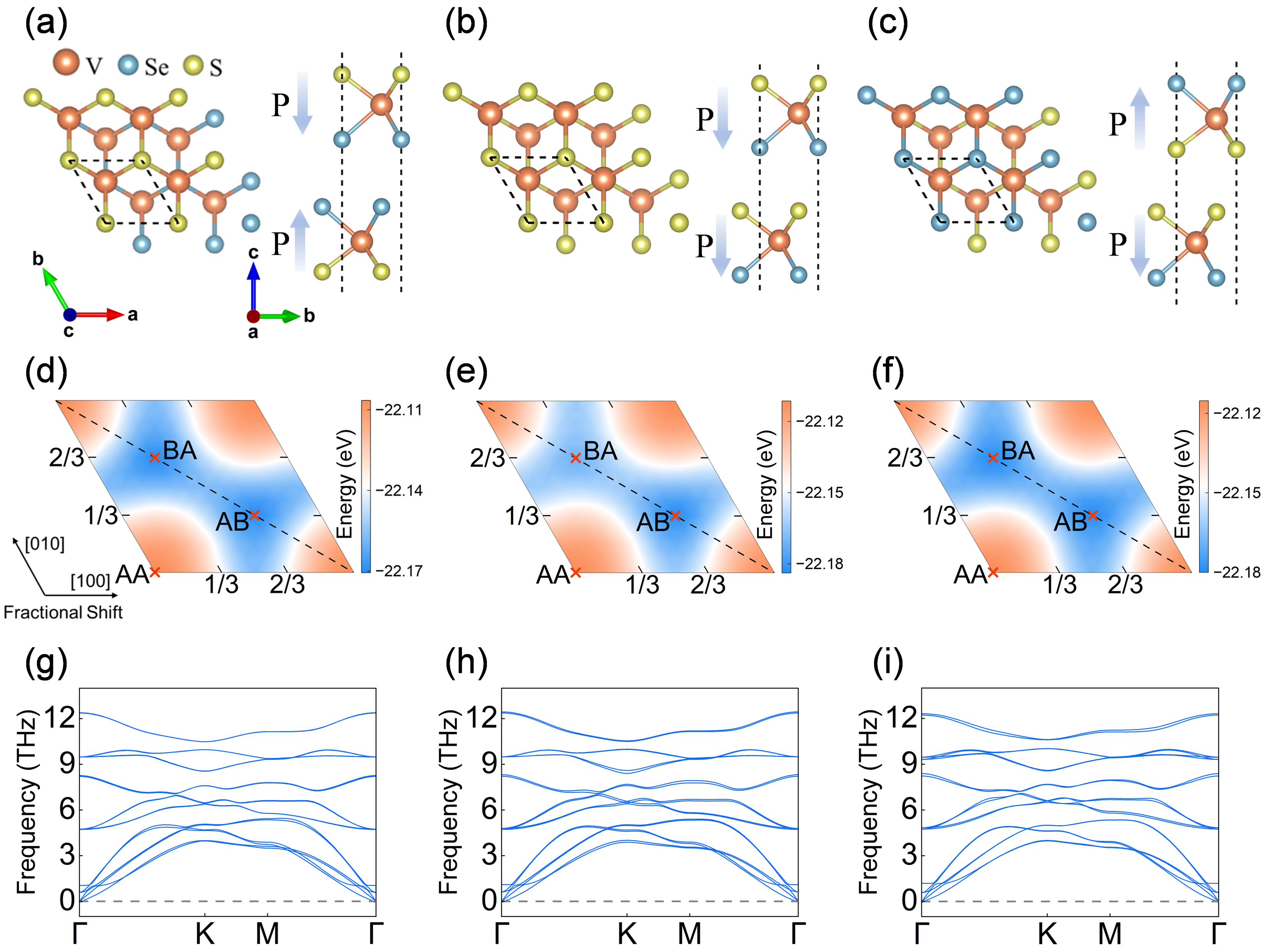}
	\caption{Top (left panel) and side (right panel) views of VSSe bilayers with (a) Se-Se, (b) Se-S, and (c) S-S interfaces. The unit cells are outlined by dashed lines. (d)--(f) The sliding energy landscapes under A-AFM state for (d) Se-Se, (e) Se-S, and (f) S-S interfaces. (g)--(i) Phonon spectra for VSSe bilayers with (g) Se-Se, (h) Se-S, and (i) S-S interfaces.}
	\label{FIG1}
\end{figure*}
	
The $2H$-$\mathrm{VSe_2}$ monolayer crystallizes in a hexagonal lattice with $D_{3h}$ symmetry, where the V atoms reside in trigonal prismatic coordination environments. It forms a sandwiched structure with a sequence of Se-V-Se, possessing a mirror plane passing through the V atomic layer. By substituting one Se atomic layer with S atoms, a Janus VSSe monolayer is constructed, in which the original mirror symmetry is broken. When two Janus VSSe monolayers are stacked to form a VSSe bilayer, distinct interfacial configurations can be obtained. Specifically, the interface of the bilayer can be formed by Se atomic layers from both the upper and lower layers [Se-Se interface, Fig.~1(a)], by Se atomic layer from the upper layer and S atomic layer from the lower layer [Se-S interface, Fig.~1(b)], or by S atomic layers from both the upper and lower layers [S-S interface, Fig.~1(c)].

To identify the energetically most favorable stacking configurations for Janus VSSe bilayers with different interfaces, we calculate the sliding energy landscape starting from the AA stacking, for which each atom of the upper layer aligns precisely with its counterpart in the lower layer. Here we only consider the case when the two constituent monolayers have parallel orientations in the basal plane, corresponding to the $R$-type stacking \cite{20-2Ren-PRL-2020}. The stacking energies were evaluated by rigidly shifting the upper layer relative to the lower layer laterally. The sliding energy landscapes for the Se-Se, Se-S, and S-S interfaces are shown in Figs.~1(d)--1(f), respectively. We can see that for Se-S interface, the system has lowest energy with a lateral shift of $(2/3,1/3)$, meaning that the upper layer sliding by $\frac{2}{3}\vec{a}+\frac{1}{3}\vec{b}$ with respect to the lower layer, termed as AB stacking. For both Se-Se and S-S interfaces, their sliding energy landscapes contain two energetically degenerate minima, corresponding to AB and BA stackings. For BA stacking, the upper layer is shifted by $\frac{1}{3}\vec{a}+\frac{2}{3}\vec{b}$ with respect to the lower layer. It should be mentioned that the above energy landscapes are calculated based on A-AFM state, that is, intralayer FM and interlayer AFM coupling. The same results can be obtained if the interlayer V atoms are ferromagnetically coupled, as shown in Fig.~S1. Since AB stacking is the common most favorable configuration for each interface, in the following, the magnetic and valley properties are investigated based on AB stackings, which are displayed in Figs.~1(a)--1(c) for Se-Se, Se-S, and S-S interfaces, respectively. Their dynamic stabilities are examined by calculating the phonon dispersions, as shown in Figs.~1(g)--1(i), respectively. No imaginary frequencies are observed, confirming the dynamic stability for all the three interfaces.	
	
\subsection{Electronic, magnetic, and valley properties}

	\begin{table*}
		\caption{Intralayer and interlayer exchange parameters (in unit of meV) for VSSe bilayers with different interfaces.}
		\label{tab:Jij}
		\begin{ruledtabular}
			\begin{tabular}{c cccc cccc}
				
				& \multicolumn{4}{c}{Intralayer exchange parameters}
				& \multicolumn{4}{c}{Interlayer exchange parameters}\\
				
				\cline{2-5}\cline{6-9}
				
				Interface
				& $J_1$ & $J_2$ & $J_3$ & $J_4$
				& $J_1^{\prime}$ & $J_2^{\prime}$ & $J_3^{\prime}$ & $J_4^{\prime}$\\
				
				\hline
				
				Se-Se
				& 19.058 & $-$1.072 & $-$0.030 & $-$0.169
				& $-$0.028 & $-$0.004 & $-$0.080 & $-$0.002 \\
				
				Se-S
				& 19.073 & $-$1.644 & 0.227 & $-$0.252
				& $-$0.006 & $-$0.007 & $-$0.007 & $-$0.0001 \\
				
				S-S
				& 17.886 & $-$2.592 & 0.524 & $-$0.194
				& $-$0.101 & 0.581 & $-$0.122 & 0.045 \\
				
			\end{tabular}
		\end{ruledtabular}
	\end{table*}

To further determine the magnetic ground state of VSSe bilayers with different interfaces, a series of possible magnetic configurations were considered, as shown in Fig.~S2. The total energies of the different magnetic configurations for each interface are summarized in Fig.~2(a). We can see that both the Se-Se and Se-S interfaces favor the A-AFM ordering, whereas the S-S interface has a FM-FM (intralayer FM and interlayer FM) ground state. The intralayer magnetic couplings are all the same for the three interfaces, and the difference comes from the interlayer coupling. We estimate the interlayer exchange interactions by calculating $E_{\mathrm{inter}} = E_{\mathrm{A\text{-}AFM}}-E_{\mathrm{FM\text{-}FM}}$, where $E_{\mathrm{FM\text{-}FM}}$ and $E_{\mathrm{A\text{-}AFM}}$ represent the total energies of the FM-FM and A-AFM states, respectively. The calculated results are $-1.12$, $-0.11$, and $0.95$~$\mathrm{meV/f.u.}$ for Se-Se, Se-S, and S-S interfaces, respectively, suggesting that the interlayer exchange couplings in VSSe bilayers are generally weak for all the interfaces.

The magnetic exchange parameters were then calculated using the Green's function method \cite{41-2green} and the results are listed in Table I. The intralayer nearest-neighbor exchange interactions ($J_1$) are positive with large magnitudes for all the interfaces, dominating in determining the robust intralayer FM coupling. In contrast, the intralayer next-nearest-neighbor exchange parameters ($J_2$) are negative, reflecting weak AFM coupling. Given that the V-X-V (X = Se, S) bond angles are close to $90^\circ$, all three interfacial configurations favor intralayer FM coupling based on the Goodenough--Kanamori--Anderson (GKA) rules \cite{46-1GKA1,46-2GKA2,46-3GKA3}. For the interlayer exchange interactions, however, the exchange parameters up to fourth-nearest neighbors are all negative for Se-Se and Se-S interfaces, consistent with the A-AFM ground states. In the case of S-S interface, however, the interlayer nearest-neighbor exchange constant ($J_1^{\prime}$) is a small negative value, in contrast to the relatively large positive value of the next-nearest-neighbor exchange constant ($J_2^{\prime}$). This indicates that the interlayer next-nearest-neighbor exchange interaction dominates in determining the interlayer FM coupling for the S-S interface.

\begin{figure*}[t]
	\centering
	\includegraphics[width=1.6\columnwidth]{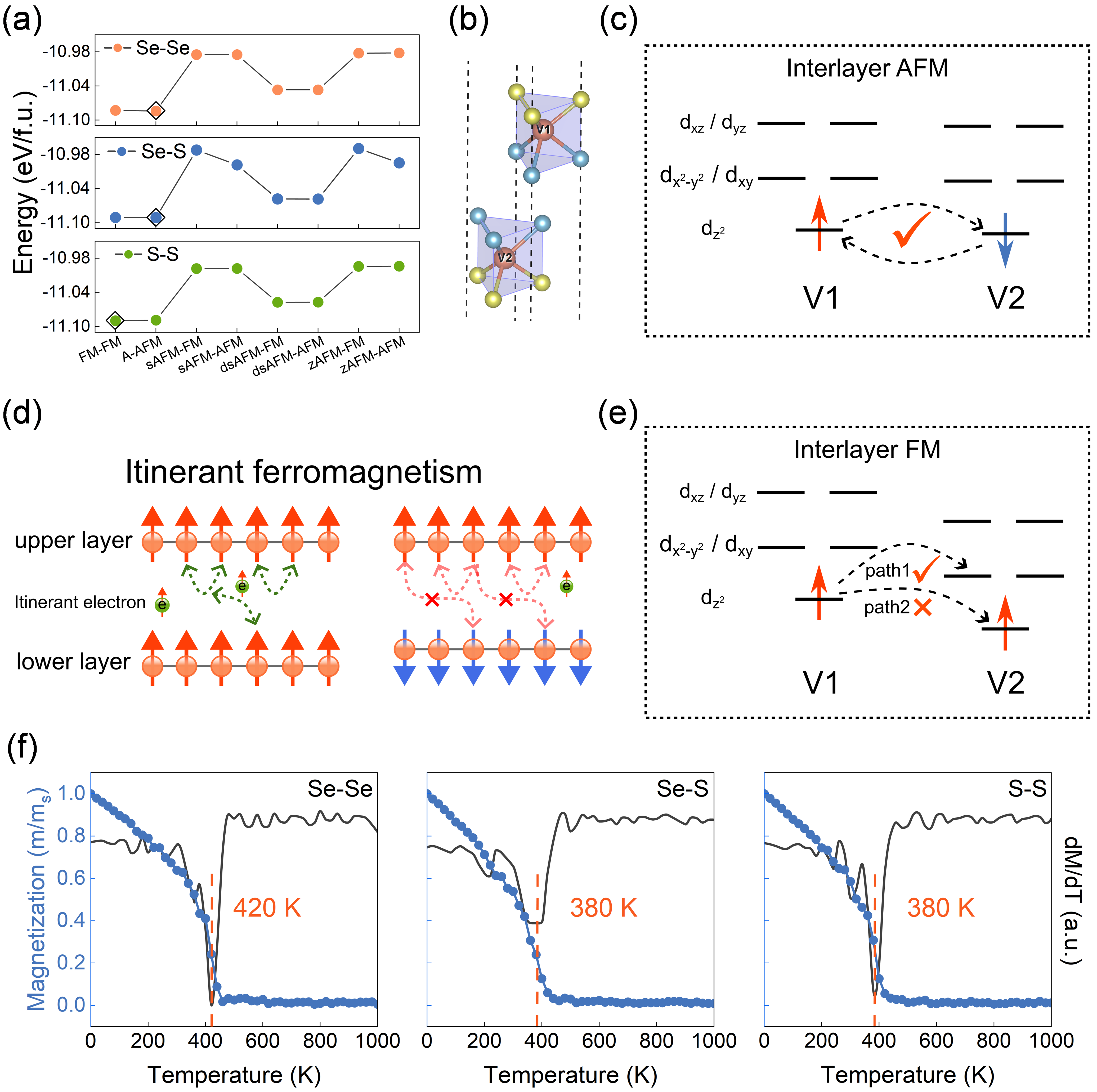}
	\caption{(a) Total energies of different magnetic configurations of VSSe bilayers with Se-Se, Se-S, and S-S interfaces. (b) Trigonal prismatic environments formed by the nearest S and Se atoms. Interlayer orbital alignment of $\mathrm{V1}$ atom from the upper layer and $\mathrm{V2}$ atom from the lower layer when (c) interlayer electrostatic potential difference $\Delta\phi$ is small and (e) $\Delta\phi$ is comparable to the energy of crystal-field splitting. Up and down arrows denote spin-up and spin-down electrons, respectively. (d) Schematic illustration of itinerant ferromagnetism induced by interlayer electron hopping in a metallic system. (f) Magnetic transition temperatures of VSSe bilayers with Se-Se, Se-S, and S-S interfaces.
	}
	\label{FIG2}
\end{figure*}

We further analyze the origin of the distinct interlayer magnetic couplings in VSSe bilayers with different interfaces. In the VSSe bilayer, each V atom is coordinated by three S atoms and three Se atoms, forming a trigonal prismatic crystal field [Fig.~2(b)]. As a result, the V-$d$ orbitals split into three energy levels: a single low-energy $d_{z^2}$ orbital, an intermediate doubly degenerate subset of $d_{x^2\mathrm{-}y^2}$ and $d_{xy}$ orbitals, and the highest doubly degenerate $d_{xz}$ and $d_{yz}$ orbitals, with the $d_{z^2}$ orbital being half-occupied. As mentioned above, the Janus structure breaks the mirror symmetry and thus within each layer of the VSSe bilayer, an intrinsic out-of-plane dipole moment is introduced. When the monolayers are stacked together, the dipole moments of upper and lower layers are head-to-head and tail-to-tail for Se-Se [Fig. 1(a)] and S-S [Fig. 1(c)] interfaces, respectively, while they are head-to-tail for Se-S interface [Fig. 1(b)]. For the Se-Se and S-S interfaces, the contribution of the electric dipoles from each constituent monolayers cancel each other and the out-of-plane electrostatic potential differences ($\Delta\phi$) mainly originate from the $R$-stacked pattern of the upper and lower layers \cite{20-2Ren-PRL-2020}. While for the Se-S interface, the dipole moments of each constituent monolayer both contribute to the $\Delta\phi$. The calculated $\Delta\phi$ are 56.6 and 4.6 meV for Se-Se and S-S interfaces, respectively, much smaller than that of the Se-S interface (763.9 meV) (Fig. S3). The built-in layer-dependent electrostatic potential consequently leads to a staggered orbital alignment across the adjacent layers. The different interlayer magnetic interactions for VSSe bilayer with different interfaces are determined by the competition between distinct interlayer hopping channels.
	
	\begin{figure*}[t]
	\centering
	\includegraphics[width=1.7\columnwidth]{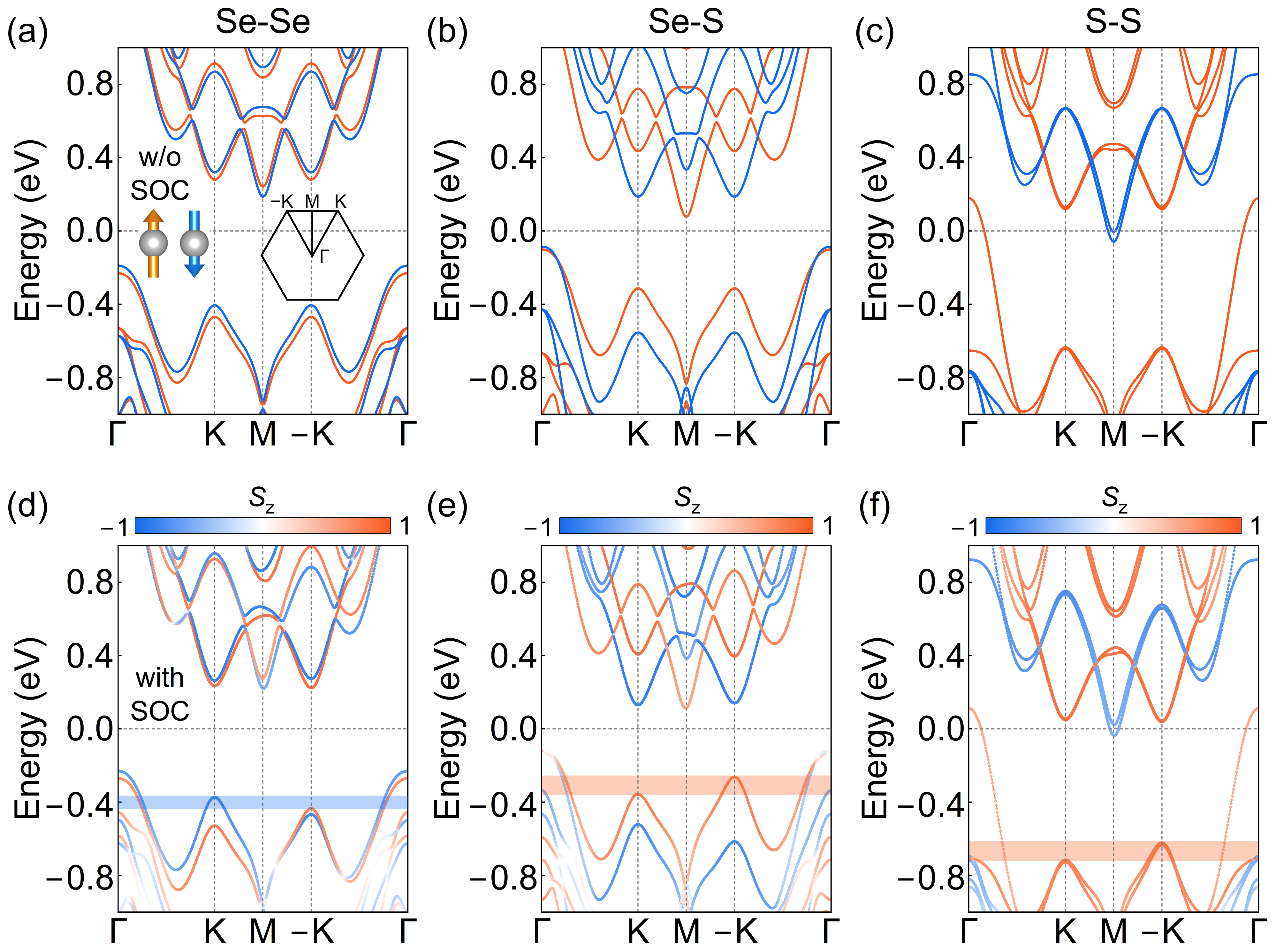}
	\caption{
		Band structures of VSSe bilayers with (a) Se-Se, (b) Se-S, and (c) S-S interfaces without considering SOC. Inset in (a) is the first Brillouin zone. Band structures of VSSe bilayers with (d) Se-Se, (e) Se-S, and (f) S-S interfaces with SOC included.
	}
	\label{FIG3}
\end{figure*}

Firstly, when $\Delta\phi$ is relatively small, the relative shift of the interlayer orbital levels is negligible, as demonstrated in Fig.~2(c). In this case, hoping of the form $d_{z^2}$-$d_{z^2}$ is allowed for interlayer AFM alignment via the super-superexchange \cite{16D.Xiao-NanoLett-2018} mediated by the interfacial S/Se atoms, whereas such hoping is prohibited for FM alignment, due to the Pauli exclusion principle. Secondly, when $\Delta\phi$ is increased, the staggered alignment of interlayer orbitals is promoted as well. If $\Delta\phi$ is large enough [Fig.~2(e)], the V-$d_{z^2}$ orbital from the upper layer will move close to the V-$d_{x^2-y^2}$/$d_{xy}$ orbitals from the lower layer, allowing for the hoping form of $d_{z^2}$-$d_{x^2-y^2}$ or $d_{z^2}$-$d_{xy}$ [path 1 in Fig.~2(e)], which will result in the interlayer FM super-superexchange interaction. Thirdly, if the system changes from a semiconductor to a metal, itinerant ferromagnetism takes place. Owing to the strong interlayer electron hopping and the pronounced charge accumulation and depletion across the interface, conduction electrons can move freely between adjacent layers, as illustrated in Fig.~2(d). The local spins tend to align parallelly when the itinerant electrons move across the adjacent layers, based on the double exchange mechanism. For this case, the parallel-spin configuration is energetically favored, stabilizing the interlayer FM metallic state.
	
The band structures in Fig.~3 reveal that VSSe bilayers with Se-Se and Se-S interfaces are semiconducting, whereas the system with S-S interface is metallic. For the Se-Se interface, the calculated $\Delta\phi$ ($56.6$~$\mathrm{meV}$) is much smaller than the typical orbital splitting induced by trigonal prismatic field \cite{47PhysRevLett.30.784}. Therefore, the interlayer exchange coupling is dominated by the hopping path shown in Fig.~2(c), for which the adjacent layers are coupled antiferromagnetically. In addition, no obvious charge accumulation or depletion is observed in the vdW gap [Fig.~S3(a)], therefore, interlayer electron hopping is strongly suppressed. For VSSe bilayer with Se-S interface, the much larger $\Delta\phi$ (763.9 meV) leads to the coexistence and competition of interlayer AFM [Fig.~2(c)] and FM [Fig.~2(e)] exchange interactions. Although the magnetic ground state for Se-S interface is interlayer AFM, the energy difference between interlayer AFM and FM states is very small, on the order of $0.1$~$\mathrm{meV}$. For the S-S interface, the interlayer $\Delta\phi$ is negligible, while the metallic property [Fig. 3(c)] guarantees that the itinerant ferromagnetism dominates in determining the interlayer FM ground state.

Magnetic anisotropy energy (MAE) plays a critical role in stabilizing the long-range magnetic ordering in 2D magnetic materials, which is calculated by rotating the magnetization direction within the $a$-$c$ and $a$-$b$ planes. The angle-dependent MAEs for the three interfacial configurations are resented in Fig.~S4. For each case, the system has the lowest energy when magnetization direction lies within the $a$-$b$ plane, and the energy difference is negligible when magnetization direction is rotated in-plane. The results indicate that the VSSe bilayers with different interfaces belong to the family of 2D $XY$ magnets, with the values of $E_{[001]}-E_{[100]}$ ($E_{[001]}$ and $E_{[100]}$ are total energies when magnetization directions are along $[001]$ and $[100]$ directions, respectively) are $618.4$, $626.5$, and $646.9$~$\mu\mathrm{eV}$ for Se-Se, Se-S, and S-S interfaces, respectively. Based on the calculated exchange parameters and MAEs, magnetic transition temperatures are further evaluated, as shown in Fig.~2(f). The Neél temperatures for VSSe bilayers with Se-Se and Se-S interfaces are $420$~$\mathrm{K}$ and $380$~$\mathrm{K}$, respectively, and the Curie temperature for S-S interface is calculated to be $380$~$\mathrm{K}$. The magnetic transition temperatures all exceed the room temperature for the three different interfaces.

We next examine the valley properties of VSSe bilayers with different interfaces. As shown in Figs.~3(a)--3(c), in the absence of SOC, the valence band maximum (VBM) for each configuration is located at the $\Gamma$ point, while the conduction band minimum (CBM) resides at the M point. For both the valence and conduction bands, the valleys at the K and $-$K points are energetically degenerate. For Se-Se and Se-S interfaces, although the magnetic ground states are both A-AFM, the breaking of inversion symmetry and built-in $\Delta\phi$ lead to the spontaneous spin splitting even without SOC. The spin splitting in Se-S interfacial configuration is much larger than that in Se-Se interface, due to the much larger $\Delta\phi$ in Se-S interface. When SOC is included, the energy degeneracy between the K and $-$K valleys is lifted, especially for the top valence bands. As a result, spontaneous valley polarization ($P_{v}$) emerges for each interfacial configuration, which is defined as $P_{v}=E_{v}^{\mathrm{K}}-E_{v}^{-\mathrm{K}}$, where $E_{v}^{\mathrm{K}}$ and $E_{v}^{-\mathrm{K}}$ are the energies of K and $-$K valleys, respectively. Since the valley polarizations of top valence bands are much larger than those of the bottom conduction bands, in the following, we focus our investigations on the former, as shown in Figs.~3(d)--3(f). The values of $P_{v}$ are $63$~$\mathrm{meV}$ for the Se-Se interface, and $94$~$\mathrm{meV}$ for both Se-S and S-S interfaces, larger than those reported in $\mathrm{CeClI}$ monolayer ($55$~$\mathrm{meV}$) \cite{48CeClI}, GdCl$_{2}$ bilayer ($46$~$\mathrm{meV}$) \cite{49GdCl2-bilayer}, and GdF$_{2}$/Sc$_{2}$CO$_{2}$ heterostructure ($41$~$\mathrm{meV}$) \cite{50GdF2-Sc2CO2}.
	
\subsection{Electric-field effect}
	
\subsubsection{Electric-field effect on electronic and magnetic properties}

	\begin{figure*}[t]
	\centering
	\includegraphics[width=1.7\columnwidth]{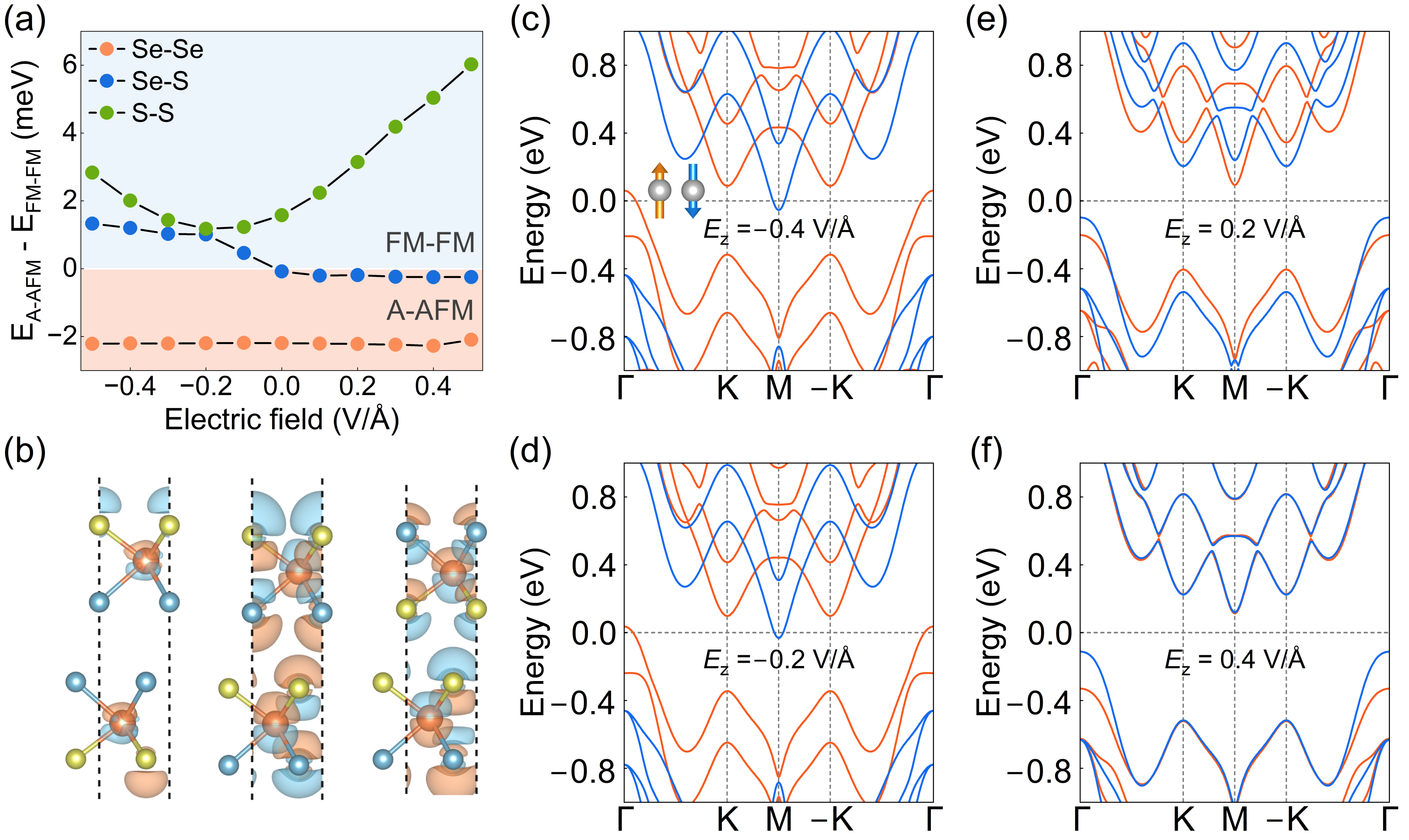}
	\caption{(a) Energy differences ($\Delta E$) between A-AFM and FM-FM states ($\Delta E=E_{\mathrm{A\text{-}AFM}}-E_{\mathrm{FM\text{-}FM}}$) as a function of out-of-plane electric field ($E_z$) for VSSe bilayers with Se-Se, Se-S, and S-S interfaces. (b) Electric-field-induced differential charge density $\Delta\rho$ defined as $\Delta\rho=\rho(E_z=-0.4)-\rho(E_z=0)$, where $\rho(E_z=-0.4)$ and $\rho(E_z=0)$ are the charge densities with $E_z=-0.4$~$\mathrm{V/\AA}$ applied and without electric field, respectively, for Se-Se (left panel), Se-S (middle panel), and S-S (right panel) interfaces. Orange and blue colors represent charge accumulation and depletion, respectively. Band structures of VSSe bilayer with Se-S interface when $E_z$ of (c) $-0.4$~$\mathrm{V/\AA}$, (d) $-0.2$~$\mathrm{V/\AA}$, (e) $0.2$~$\mathrm{V/\AA}$, and (f) $0.4$~$\mathrm{V/\AA}$ are applied, without considering SOC. The orange and blue lines represent spin-up and spin-down channels, respectively.
	}
	\label{FIG4}
   \end{figure*}

The out-of-plane electric fields ($E_z$) ranging from $-0.5$ to $0.5$~$\mathrm{V/\AA}$ are applied to the VSSe bilayers with different interfaces. The energy differences ($\Delta E$) between A-AFM and FM-FM states ($\Delta E=E_{\mathrm{A\text{-}AFM}}-E_{\mathrm{FM\text{-}FM}}$) as a function of $E_z$ are shown in Fig.~4(a). We can see that the magnetic ground states are not changed by the electric fields for VSSe bilayers with Se-Se and S-S interfaces. In contrast, for the Se-S interface, the interlayer magnetic coupling can be effectively tuned by $E_z$. Under a negative $E_z$, the magnetic ground state changes from interlayer AFM to FM ordering, and $\Delta E$ increases monotonically with increasing the magnitude of $E_z$. However, under a positive $E_z$, $\Delta E$ remains negative and the A-AFM state is further stabilized by the electric field. 
	
To investigate the origin of the different electric-filed effects on the magnetic ground states, we calculate the differential charge density, defined as $\Delta\rho=\rho_{\mathrm{bilayer}}-\rho_{\mathrm{upper\text{-}layer}}-\rho_{\mathrm{lower\text{-}layer}}$, where $\rho_{\mathrm{bilayer}}$ is the charge density of VSSe bilayer, and $\rho_{\mathrm{upper\text{-}layer}}$ and $\rho_{\mathrm{lower\text{-}layer}}$ are the charge densities of the upper and lower layers of VSSe bilayer, respectively, before the bilayer is formed. The results for VSSe bilayers with different interfaces under a series of electric fields are shown in Fig.~S5. For the Se-Se interface, the charge redistribution in the vdW gap remains weak throughout the entire electric-field range, indicating the weak interlayer interaction. This is further supported by the field-induced charge density difference defined as $\Delta\rho=\rho(E_z)-\rho(0)$, where $\rho(E_z)$ and $\rho(0)$ are the charge densities after and before the electric filed $E_z$ is applied, respectively. The result for Se-Se interface under $E_z=-0.4$~$\mathrm{V/\AA}$ is shown in the left panel of Fig.~4(b). No obvious charge accumulation or depletion is observed in the vdW gap, even under a relatively strong electric field. The system remains semiconducting for both positive and negative electric fields, as shown in Fig.~S6. Consequently, interlayer electron hopping can’t be triggered by external electric field, preventing the establishment of effective itinerant-electron-mediated FM exchange paths. Under such condition, the interlayer magnetic coupling is still dominated by the AFM super-superexchange interactions. Therefore, the applied electric field is insufficient to induce a magnetic phase transition, and the system robustly preserves the A-AFM ground state.

	\begin{figure*}[t]
		\centering
		\includegraphics[width=1.7\columnwidth]{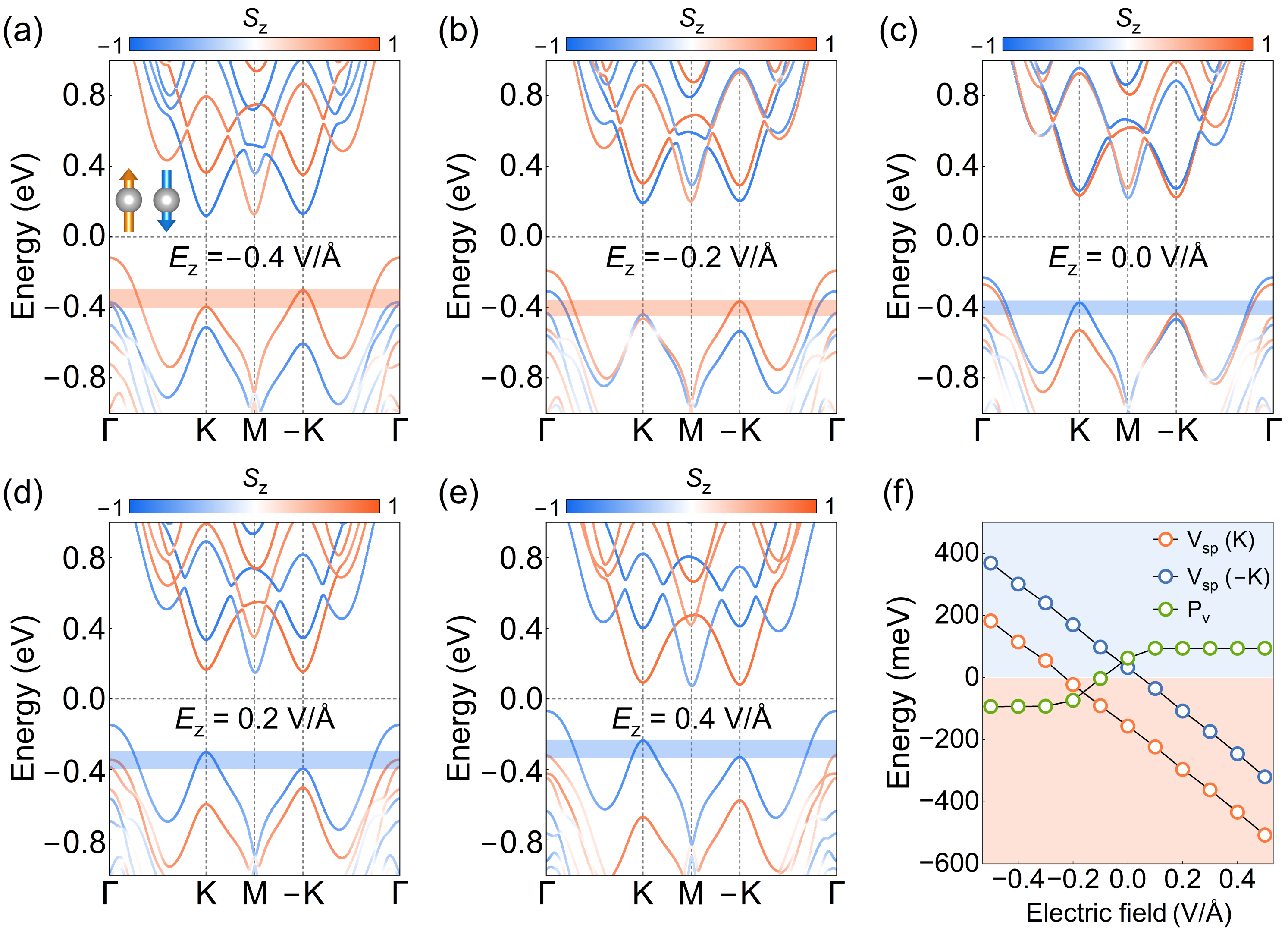}
		\caption{Spin-resolved band structures of VSSe bilayers with Se-Se interface under out-of-plane electric fields ($E_z$) of (a) $-0.4$~$\mathrm{V/\AA}$, (b) $-0.2$~$\mathrm{V/\AA}$, (c) $0.0$~$\mathrm{V/\AA}$, (d) $0.2$~$\mathrm{V/\AA}$, and (e) $0.4$~$\mathrm{V/\AA}$ with SOC included. Orange and blue lines represent spin-up and spin-down channels, respectively. (f) Valley polarization ($P_{v}$) of K and $-$K valleys and the corresponding spin splitting ($V_{\mathrm{sp}}$) at K and $-$K valleys as a function of $E_z$.
		}
		\label{FIG5}
	\end{figure*}

In contrast, the electric-field-induced magnetic phase transition emerges in VSSe bilayer with the Se-S interface. To be specific, a transition from interlayer AFM to FM state occurs when a negative $E_z$ is applied, while the system remains in the interlayer AFM ground state under a positive $E_z$. The calculated differential charge densities [Fig.~S5(b)] show that the charge accumulation in the vdW gap is significantly enhanced under negative $E_z$ but weakened under positive $E_z$. This is further demonstrated by the field-induced charge density difference shown in the middle panel of Fig.~4(b), in which pronounced charge redistribution is observed when a negative $E_z$ is applied. Therefore, the negative $E_z$ facilitates the interlayer electron hopping. As the charge redistribution increases, the interlayer itinerant ferromagnetism is gradually enhanced. Once the interlayer electron hopping is sufficiently strong, the interlayer itinerant FM interaction exceeds the competing AFM/FM super-superexchange interactions discussed above, driving an interlayer AFM-to-FM phase transition, accompanied by a semiconductor-to-metal transition [Figs.~4(c) and 4(d)]. Conversely, under a positive $E_z$, the reduced charge accumulation in the vdW gap suppresses the interlayer electron hopping and weakens the itinerant-electron contribution of the FM exchange interaction. As a result, the system remains its AFM semiconducting state [Figs.~4(e) and 4(f)]. These results indicate that the VSSe bilayer with Se-S interface is located near the critical boundary between competing interlayer AFM and FM states, rendering its magnetic ground state highly sensitive to external stimuli.
	
For the S-S interface, the system remains in a robust FM metallic state throughout the entire electric-field range of $-0.5$ to $0.5$~$\mathrm{V/\AA}$ [Fig.~4(a) and Fig.~S7]. The differential charge density [Fig.~S5(c)] exhibits charge accumulation in the vdW gap even without electric field, indicating intrinsically strong electron delocalization. Furthermore, for both positive and negative $E_z$, the interlayer charge redistribution continuously increases with increasing $E_z$ [Fig.~S5(c) and right panel of Fig.~4(b)], suggesting that the external $E_z$ further enhances the interlayer electron hopping, stabilizing the interlayer FM exchange coupling, which is also corroborated by the electric-field-dependent energy difference between A-AFM and FM-FM states displayed in Fig.~4(a).
	
	\begin{figure*}[t]
	\centering
	\includegraphics[width=1.6\columnwidth]{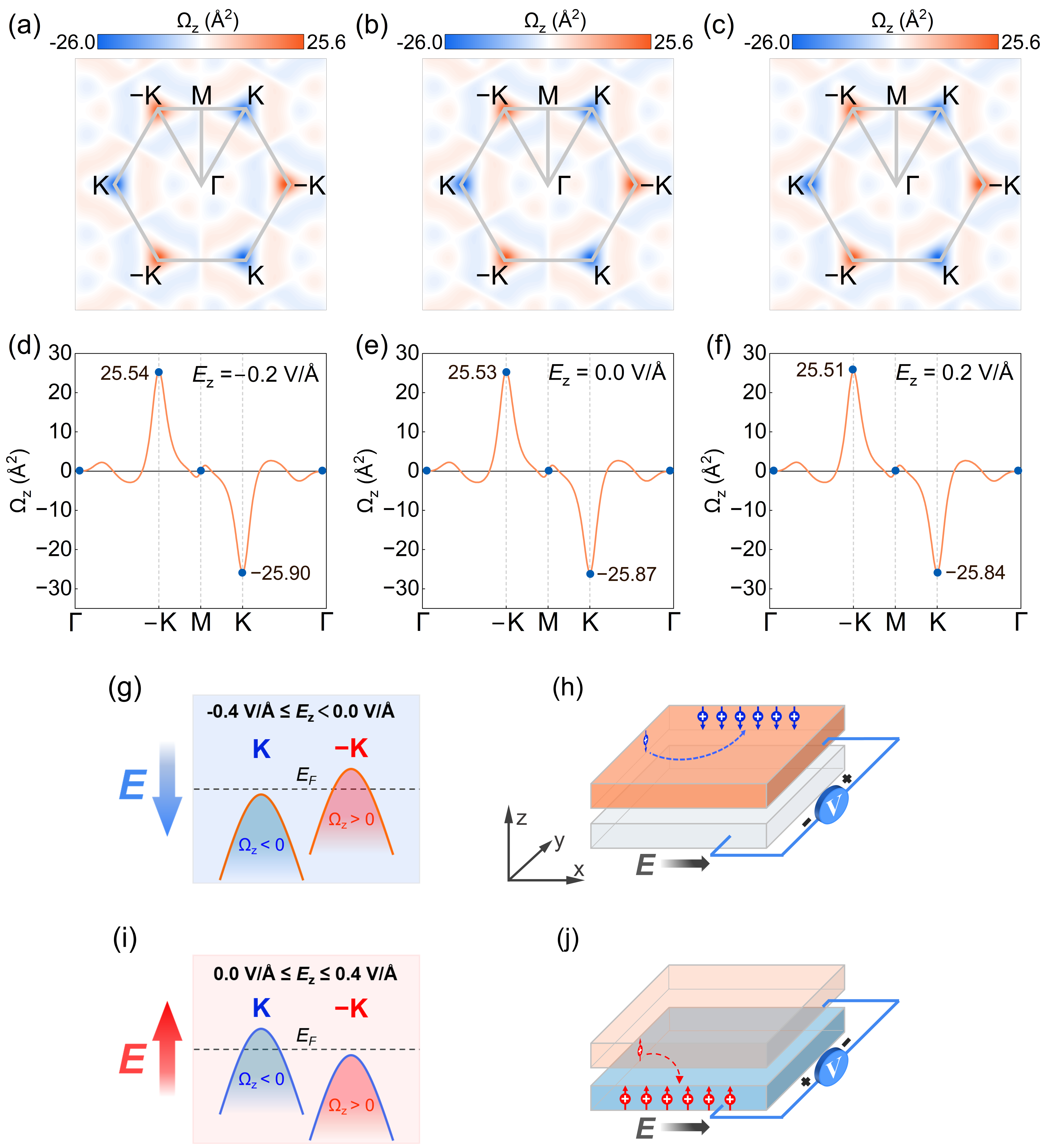}
	\caption{Berry curvature distributions across the Brillouin zone when out-of-plane electric fields $E_z$ of (a) $-0.2$~$\mathrm{V/\AA}$, (b) $0.0$~$\mathrm{V/\AA}$, and (c) $0.2$~$\mathrm{V/\AA}$ are applied. Berry curvatures along high-symmetry directions when $E_z$ of (d) $-0.2$~$\mathrm{V/\AA}$, (e) $0.0$~$\mathrm{V/\AA}$, and (f) $0.2$~$\mathrm{V/\AA}$ are applied. Schematic illustration of valley polarization and signs of Berry curvatures at K and $-$K valleys for (g) negative and (i) positive $E_z$. Schematic diagram of an AVH device under (h) negative and (j) positive $E_z$.
	}
	\label{FIG6}
\end{figure*}
	
\subsubsection{Electric-field effect on valley polarization}
	
Next, we investigate the effect of $E_z$ on spin and valley properties of the VSSe bilayers when SOC is considered, with the results for Se-Se interface shown in Fig.~5. The spin splitting at a particular valley is defined as $V_{\mathrm{sp}}=E_{v}^{\uparrow}-E_{v}^{\downarrow}$, where $E_{v}^{\uparrow}$ and $E_{v}^{\downarrow}$ represent energies of the spin-up and spin-down bands, respectively. When no electric field is applied, the valley polarization $P_{v}$ is calculated to be $63$~$\mathrm{meV}$ for Se-Se interface, and the spin splittings at K and $-$K valleys are $-157$ and $32$~$\mathrm{meV}$, respectively. Layer-resolved band structure [Fig.~S8(c)] reveals that the top valence band at K ($-$K) valley is contributed by the V atoms from lower (upper) layer, exhibiting a spin-valley-layer locked property. Under a negative $E_z$ of $-0.2$~$\mathrm{V/\AA}$, the energy of the upper layer is elevated, while that of the lower layer is relatively lowered [Fig.~S8(b)]. Consequently, the energy of the $-$K valley tends to be larger than that of the K valley, and the valley polarization is reversed [Fig.~5(b)]. When $E_z$ is increased to be $-0.4$~$\mathrm{V/\AA}$, the energy of spin-up band exceeds that of the spin-down band at K valley [Fig.~5(a)], and both the K and $-$K valleys are contributed by V atoms from the upper layer [Fig.~S8(a)], that’s to say, the spin-valley-layer locking is broken. As shown in Fig.~5(f), the valley polarization is gradually reversed and increased by the negative $E_z$ and eventually saturates at approximately $-94$~$\mathrm{meV}$. The spin splitting at $-$K valley increases monotonically as a function of the negative $E_z$, while for the K valley, it firstly decreases with increasing the negative $E_z$, and then changes its sign and increases monotonically when $E_z$ is further increased.

In contrast, a positive $E_z$ raises the energy of the lower layer and suppresses the energy of the upper layer. For positive $E_z$, the energy of spin-down band contributed by V atoms from the lower layer exceeds that of the spin-up band contributed by V atoms from the upper layer. Therefore, both K and $-$K valleys are dominated by the spin-down bands contributed by V atoms from the lower layer [Figs.~5(d) and 5(e); Figs.~S8(d) and S8(e)], whereas the sign of valley polarization is unchanged by the electric field. As shown in Fig. 5(f), the valley polarization is increased to be 94 meV for a positive $E_z$ of $0.1$~$\mathrm{V/\AA}$ and becomes saturated when $E_z$ is further increased. The spin splitting at K point keeps negative and the magnitude increases monotonically as a function of $E_z$. For spin splitting at $-$K valley, however, it firstly changes its sign when a positive $E_z$ is applied, and then increases monotonically when $E_z$ is further increased.
	
The VSSe bilayers with Se-S and S-S interfaces exhibit remarkably different responses to the applied electric field, as shown in Figs.~S9 and S10, respectively. Although SOC also induces valley polarization in these two configurations, the energy of the K valley keeps lower than that of the $-$K valley throughout the range of $E_z$ we considered, regardless of the direction of $E_z$. Only the magnitude of valley polarization can be modulated by the out-of-plane electric field, but the sign cannot be reversed. For Se-S interface [Fig.~S9], the large value of the built-in $\Delta\phi$ induces a relatively large spin splitting even through the system is in the A-AFM ground state, and both K and $-$K valleys are contributed by V atoms from the upper layer with the spin-up channels [Fig. S9(c) and S9(h)]. When a negative $E_z$ is applied, the system tends to be a FM metal. For a positive $E_z$, however, the system keeps semiconducting, and the energy of the lower layer is gradually increased by the electric field, and at $E_z$ of $0.4$~$\mathrm{V/\AA}$ , the spin splitting at K valley is reversed. For the S-S interface, the system remains a FM metal for both positive and negative $E_z$, and the signs of valley polarization as well as spin splitting keep unchanged.

\subsubsection{All-electric-field controlled anomalous valley Hall effect}
	
Since both valley polarization and spin splitting can be reversed in VSSe bilayer with Se-Se interface, in the following, we focus on this configuration to investigate the anomalous valley Hall (AVH) effect. The intrinsic anomalous Hall conductivity is often related to the Berry curvature of the occupied bands \cite{51-1Berry,52RevModPhys.82.1539}. Under the Kubo formula \cite{51-1Berry}, the Berry curvature in the $z$ direction can be derived as
	\[\Omega_z(\mathbf{k})=-\sum_{n}\sum_{m\neq n}f_{n\mathbf{k}}\frac{
		2\,\mathrm{Im}
		\langle \psi_{n\mathbf{k}}|\hat{v}_x|\psi_{m\mathbf{k}}\rangle
		\langle \psi_{m\mathbf{k}}|\hat{v}_y|\psi_{n\mathbf{k}}\rangle
	}
	{(E_{m\mathbf{k}}-E_{n\mathbf{k}})^2},
	\]
where $f_{n\mathbf{k}}$ is the Fermi-Dirac distribution function, $\psi_{n\mathbf{k}}$ ($\psi_{m\mathbf{k}}$) is the Bloch wave function with eigenenergy $E_{n\mathbf{k}}$ ($E_{m\mathbf{k}}$), and $\hat{v}_x$ and $\hat{v}_y$ are the velocity operators along the $x$ and $y$ directions, respectively. For nonmagnetic transition metal dichalcogenides, MoS$_2$ for example, the Berry curvature has an odd parity $\Omega_z(\mathrm{K})=-\Omega_z(-\mathrm{K})$ due to the time-reversal symmetry \cite{53PhysRevB.77.235406,54PhysRevB.92.125146}. For the VSSe bilayer with Se-Se interface, the signs of $\Omega_z$ near the K and $-$K points are also opposite, but the magnitudes are different [Figs.~6(b) and 6(e)]. The signs of $\Omega_z(\mathrm{K})$ and $\Omega_z(-\mathrm{K})$ are not changed by the applied out-of-plane electric field $E_z$, as shown in Figs. 6(a) and 6(d) for $E_z=-0.2~\mathrm{V/\AA}$ and in Figs. 6(c) and 6(f) for $E_z=0.2~\mathrm{V/\AA}$. However, under a negative $E_z$, the energy of the $-$K valley becomes higher than that of the K valley [Fig.~6(g)]. With proper hole doping, the Fermi energy ($E_\mathrm{F}$) can be moved between K and $-$K valleys, and the majority carriers are spin-down ones predominantly contributed by $-$K valley, which has a positive $\Omega_z$. Under an in-plane longitudinal electric field $\vec{E}$, the carriers acquire an anomalous transverse velocity $\vec{v}$, based on $\vec{v}\sim-\vec{E}\times\Omega_z(\mathbf{k})$. In this case, the spin-down majority carriers coming from the $-$K valley dominate the transverse transport. Since the $-$K valley is primarily contributed by the atoms from upper layer [Figs.~S8(a) and S8(b)], these carriers acquire an anomalous velocity $\vec{v}$ along $y$ direction if the in-plane electric field $\vec{E}$ is applied along $x$ direction, and accumulate at the edge of the upper layer, generating a charge Hall current that can be detected as a negative voltage, as illustrated in Fig.~6(h). It is noted that the net charge comes from the same valley with the same spin, resulting in a simultaneous charge, spin, and valley polarization, termed as AVH effect. In contrast, when a positive $E_z$ is applied, the valley polarization is reversed, and the energy of K valley is higher than that of $-$K valley [Fig.~6(i)]. Since K valley has a nagative $\Omega_z$ and is mainly contributed by the atoms from the lower layer [Fig.~S8(d) and S8(e)], the spin-up majority carriers from the K valley accumulate at the opposite edge of the lower layer under the same in-plane electric field, and the AVH voltage is reversed [Fig.~6(j)]. Therefore, in VSSe bilayer with Se-Se interface, the AVH voltage can be effectively reversed by changing the direction of the out-of-plane electric field $E_z$, realizing an all-electric-field controlled AVH effect.

\section{Conclusion}
	
In summary, we predict by first-principles calculations that Janus VSSe bilayers with different interfaces possess distinct interlayer magnetic couplings and show dramatically varied responses to external electric fields. While the magnitudes of dipole moments within each constituent monolayer are identical, the interlayer $\Delta\phi$ depends strongly on the interfacial configurations. For Se-S interface, a negative out-of-plane electric field facilitates the interlayer electron hopping and the system changes from a semiconductor to a metal, accompanied by an interlayer AFM to FM transition. The reversible magnetic phase transition can be understood by the competition between itinerant-electron-mediated FM exchange interaction and interlayer $\Delta\phi$-dependent AFM/FM super-superexchange interactions mediated by interfacial Se and S atoms. In addition, both the valley polarization and spin splitting in VSSe bilayer with Se-Se interface can be effectively reversed by the out-of-plane electric field, realizing an all-electric-field controlled AVH effect. These results identify Janus VSSe bilayer a promising platform for the design of low-power spintronic and valleytronic devices.
	
\section{Acknowledgement}
	
This work was supported by the National Natural Science Foundation of China (Grant Nos.~12474112 and 12174080) and the National Key R\&D Program of China (Grant 
No.~2022YFA1602601). The calculations were completed on the HPC Platform of Hefei University of Technology.

\end{document}